\documentclass{article} 
\usepackage{epsf}

\begin{document}
\newcommand{\braket}[2]{\langle #1 | #2 \rangle}
\newcommand{\ket}[1]{\left | \, #1 \right \rangle}
\newcommand{\mod}{\mbox{\rm \/mod\ }}
\newcommand{\bra}[1]{\left \langle #1 \, \right |}
\newcommand{\proj}[1]{\ket{#1}\!\!\bra{#1}}

\title{On the Number of Elements Needed
in a POVM Attaining the Accessible Information}

\author{Peter W. Shor\\
AT\&T Labs---Research\\
Florham Park, NJ 07932, USA\\
shor@research.att.com}

\maketitle


\begin{abstract}
We investigate an symmetric set of three quantum states in three 
dimensions with interesting properties, 
which we call the lifted trine states.  We show that for
the ensemble consisting of the three lifted trine states taken with
equal probabilities, the POVM measurement realizing the accessible 
information must contain six projectors, giving a counter-example to
a conjecture of Levitin.
\end{abstract}

Accessible information was one
of the first information-theoretic 
quantities investigated with respect to quantum systems.  
The accessible information of
an ensemble of quantum states is the maximum mutual information obtainable
between the states of the ensemble and the outcomes of a POVM (positive 
operator valued measurement) on these states.
In this paper, we investigate
how complicated a measurement which
achieves the accessible information must be.  
Davies' theorem gives a
maximum on the number of elements of the POVM needed to attain the accessible
information; namely, 
if the ensemble being considered is contained in a
$d$-dimensional Hilbert space, then at most $d^2$ elements are needed in
an optimal POVM.  When all the states are real, this bound can be improved
to $d(d-1)/2$ \cite{Davies-gen}.  C. Fuchs and A. Peres~\cite{Fuchs-pc} 
have done numerical studies on ensembles containing 
only two elements.  They found no examples where more than $d$ states 
were needed; that is, they found that the optimal measurement could always 
be a von Neumann measurement.  In two dimensions, this was proved by 
Levitin~\cite{Levitin-conj}, who also conjectured that 
in $d$ dimensions, if the number of quantum states in the ensemble 
is at most $d$, a von Neumann measurement is sufficient to attain
the accessible information.  In this paper, we give an ensemble of 
three real quantum states in three dimensions, where a POVM attaining
the accessible information must contain at least six elements, the
maximum by Davies' theorem. 

We investigate the accessible information of an ensemble consisting
of three quantum states we call the {\em lifted trine states}, with equal
probabilities on these states.
The lifted trine states are obtained by starting with the two-dimensional
quantum trine states: $(1,0)$, $(-1/2,\sqrt{3}/2)$, $(-1/2, -\sqrt{3}/2)$,
introduced by Holevo \cite{Holevo-trines} and later studied by Peres 
and Wootters \cite{Peres-Wootters}.
We add a third dimension to the Hilbert space of the trine states,
and lift all of the trine
states out the plane into this dimension by
an angle of $\arcsin \sqrt{\alpha}$, so the states become 
$(\sqrt{1-\alpha}, 0, \sqrt{\alpha})$, and so forth.
We will be dealing with small $\alpha$ (roughly, $\alpha < .1$),
so that they are close to being
planar.   This is the most interesting regime.  When 
the trine states are lifted further out of the plane, they start
behaving in relatively uninteresting ways until they are close to being
vertical; then they start being interesting again, but this 
second regime is beyond the scope of this paper.
The lifted trine states are thus:
\begin{eqnarray}
T_0(\alpha) &=& (\sqrt{1-\alpha}, 0, \sqrt{\alpha})
\nonumber
\\
T_1(\alpha) &=& (-{\textstyle{\frac{1}{2}}}\sqrt{1-\alpha}, 
{\textstyle{\frac{\sqrt{3}}{2}}}\sqrt{1-\alpha}, \sqrt{\alpha})\\
T_2(\alpha) &=& (-{\textstyle{\frac{1}{2}}}\sqrt{1-\alpha}, 
- {\textstyle{\frac{\sqrt{3}}{2}}}\sqrt{1-\alpha}, 
\sqrt{\alpha})
\nonumber
\end{eqnarray}
When it is clear what $\alpha$ is, we may drop it from the notation
and use $T_0$, $T_1$, and $T_2$.

In this section, we find the accessible information for this ensemble
of lifted trine states.  The accessible information is defined
as the maximal
mutual information between the trine states (with probabilities $\frac{1}{3}$
each) and the elements of a POVM measuring these states.
Because the lifted trine states are real vectors, it follows from the
version of Davies' theorem for real states \cite{Davies-gen}
that there is an optimal POVM
with at most six elements, all the components of which are real.  
The lifted trine states are three-fold symmetric, so by symmetrizing we
can assume that the optimal POVM is three-fold symmetric (possibly at
the cost of introducing extra POVM elements).  Also, 
the optimal POVM can be taken to have one-dimensional elements $E$, 
so the elements
can be described as vectors $\ket{v_i}$ where $E_i = \ket{v_i} \bra{v_i}$.  
This means that there is an 
optimal POVM whose vectors come in triples of the form: 
$\sqrt{p} P_0(\phi,\theta)$,
$\sqrt{p} P_1(\phi,\theta)$,
$\sqrt{p} P_2(\phi,\theta)$,
where $p$ is a scalar probability and 
\begin{eqnarray}
\nonumber
P_0(\phi, \theta) &=& (\cos\phi\cos\theta, 
\cos\phi\sin\theta, \sin\phi)
\\
P_1(\phi, \theta) &=&  
(\cos\phi\cos(\theta + 2 \pi/3), 
\cos\phi\sin(\theta + 2 \pi/3), \sin\phi)
\\
P_2(\phi, \theta) &=& 
(\cos\phi\cos(\theta - 2 \pi/3), 
\cos\phi\sin(\theta - 2 \pi/3), \sin\phi).
\nonumber
\end{eqnarray}

Suppose that the optimal POVM has several such triples, which we call
$\sqrt{p_1}\, P_b(\phi_1, \theta_1)$, 
$\sqrt{p_2}\, P_b(\phi_2, \theta_2)$, $\ldots$,
$\sqrt{p_m}\, P_b(\phi_m, \theta_m)$.
It is easily seen that
the conditions for this set of vectors to be a POVM are that
\begin{equation}
    \sum_{i=1}^m   p_i \sin^2(\phi_i) = 1/3                    
{\mathrm{\quad and\quad }}
    \sum_{i=1}^m   p_i  = 1.
\label{constraint1}
\end{equation}
The formula for accessible information 
$I_A$ can be broken
into pieces so that each triple contributes a linear amount to $I_A$.  
That is, $I_A$ is the weighted average (weighted according 
to $p_i$) of some contribution $I(\phi,\theta)$
from each $(\phi, \theta)$.  To show this, recall that $I_A$ is the
mutual information between the input and the output, and this can be expressed
as the entropy of the input less the entropy of the input given the output,
$H(X_\mathrm{in}) -  H(X_\mathrm{in}|X_\mathrm{out})$.
The term $H(X_\mathrm{in}|X_\mathrm{out})$ naturally decomposes into terms
corresponding to the various POVM outcomes, and there are several ways of
assigning the entropy of the input
$H(X_\mathrm{in})$ 
to the various POVM elements in order to complete this 
decomposition.  Following
this analysis eventually gives the same answer as is obtained
below (and is in fact how I arrived at it).  I briefly sketch this analysis 
so as to give the intuition
behind it, and then go into detail in a second analysis, which 
is superior in that it
explains the form of the answer.

For each $\phi$, and each $\alpha$, there is a $\theta$ that optimizes
$I(\phi,\theta)$.  
This $\theta$ starts out
at $\pi/6$ for $\phi=0$, decreases until it hits 0 at some value of $\phi$
(which depends on $\alpha$), 
and stays at $0$ until $\phi$ reaches its maximum value of $\pi/2$.  
For a fixed $\alpha$, by finding (numerically)
the optimal value of $\theta$ for each $\phi$ and using it to obtain
the contribution to $I_A$ attributable to that~$\phi$, we get a curve 
giving the optimal contribution to $I_A$ for each $\phi$.  
If this curve is plotted,
with the $x$-value being $\sin^2 \phi $ and the 
$y$-value being the contribution to $I_A$,
an optimal POVM is obtained from the set of points on this curve
whose average $x$-value is $1/3$ (from Eq.~\ref{constraint1}), and whose
average $y$-value is as large as possible given this constraint on the
$x$-values.  A simple convexity
argument shows that we only need at most two points from the curve
to obtain this
optimum, and that we will need one or two points depending on whether the
relevant part of the curve is concave or convex.  
For small $\alpha$, it turns out that the relevant piece of the
curve is convex, and we need two $\phi$'s to
achieve the maximum.  Each of these $\phi$'s corresponds to a triple
of POVM elements.   One of the $(\phi,\theta)$ pairs is $(0,\pi/6)$, 
and the other is $(\phi_{\alpha},0)$ for some  
$\phi_\alpha > \arcsin(1/\sqrt{3})$.  The formula for this
$\phi_\alpha$ will be derived later.

The analysis in the remainder of this section shows that
this six-outcome optimal POVM can be described in a different way,
which unifies the optimal measurements for the different $\alpha$'s.  
For small $\alpha$ ($\alpha < \gamma_1$ for some constant~$\gamma_1$), 
we first 
take the trine $T_b(\alpha)$ and make a partial
measurement which either
projects it down to the $x,y$ plane or lifts it further out of the
plane so that it becomes
the trine $T_b(\gamma_1)$.
(Note that $\gamma_1$ is
independent of $\alpha$.)  If the trine was projected into the
$x,y$ plane, we make a second measurement using the POVM with 
outcome vectors $\sqrt{2/3}(0,1)$  and
$\sqrt{2/3} (\pm\sqrt{3}/2, -1/2)$.  This is the optimal POVM for trines
in the $x,y$-plane.  
If the trine was lifted up, we use the von
Neumann measurement that projects onto the basis containing
$(\sqrt{2/3}, 0, \sqrt{1/3})$ and 
$(-\sqrt{1/6}, \pm \sqrt{1/2}, \sqrt{1/3})$.  
If $\alpha$ 
is larger than $\gamma_1$ (but still smaller than $8/9$) 
we skip the first partial measurement, and just
use the above von Neumann 
measurement.   Here, $\gamma_1$
is obtained by numerically solving a fairly complicated equation;
we suspect that no closed form expression for $\gamma_1$ exists.  
The value of $\gamma_1$ is .061367, which is $\sin^2 \phi$ for $\phi =
.25033$ radians ($14.343^\circ$).

We now give more details on this decomposition of
the POVM into a two-step
process.  We first apply a partial measurement which does not extract
all of the quantum information, i.e., it leaves a quantum residual state
that is not completely determined by the measurement outcome.
Formally, we apply one of a set
of matrices $A_i$ satisfying $\sum_i A_i^\dagger  A_i= I$.  If we start
with a pure state $\ket{v}$, we observe the $i$'th outcome with probability
$\bra{v} A_i^\dagger A_i \ket{v}$, and in this case the state $\ket{v}$ 
is taken to the state $A_i \ket{v}$.  For our purposes, we choose as the
$A_i$'s the matrices $\sqrt{p_i}\, M(\phi_i)$ where
\begin{equation}
M(\phi) = 
\left(
\begin{array}{ccc}
\sqrt{\frac{3}{2}} \cos \phi  & 0 & 0 \\
0 &  \sqrt{\frac{3}{2}} \cos \phi & 0 \\
0 & 0 & \sqrt{3} \sin \phi 
\end{array}
\right)
\end{equation}
The $\sqrt{p_i}\,M(\phi_i)$ will form a valid partial measurement
if and only if 
$\sum_i p_i \sin^2(\phi_i)$ $=$ $1/3$ and $\sum_i p_i = 1$, the same 
conditions [Eq.~(\ref{constraint1})]
as for the $P_b(\phi_i,\theta_i)$.
By first applying the above $\sqrt{p_i}\,M(\phi_i)$, and then applying 
the von Neumann 
measurement with the three basis vectors
\begin{eqnarray}
\nonumber
V_0(\theta) &= & \textstyle \Big(\sqrt{\frac{2}{3}}\cos(\theta), 
\sqrt{\frac{2}{3}}\sin(\theta), \frac{1}{\sqrt{3}}\Big) \\
\label{defineV}
V_1(\theta) &=& \textstyle \Big(\sqrt{\frac{2}{3}}\cos(\theta+2 \pi/3 ), 
\sqrt{\frac{2}{3}}\sin(\theta+2 \pi/3), \frac{1}{\sqrt{3}}\Big) \\
V_2(\theta) &=& \textstyle \Big(\sqrt{\frac{2}{3}}\cos(\theta-2 \pi/3 ), 
\sqrt{\frac{2}{3}}\sin(\theta-2 \pi/3), \frac{1}{\sqrt{3}}\Big) 
\nonumber
\end{eqnarray}
we obtain the POVM given by
the vectors $\sqrt{p_i} \, P_b(\theta_i, \phi_i)$;  
checking this is simply a matter of verifying that 
$V_b(\theta) M(\phi) = P_b(\theta,\phi)$.  
Now, after applying $\sqrt{p_i}\, M(\phi_i)$ to the trine $T_0(\alpha)$, 
we get the vector
\begin{equation}
\big(\sqrt{3/2}\sqrt{1-\alpha}\sqrt{p_i} \cos \phi_i , 0, 
\sqrt{3} \sqrt{\alpha} \sqrt{p_i} \sin \phi_i  \big).
\end{equation}
This is just the state
$\sqrt{p_i'}\, T_0(\alpha_i')$ where $T_0(\alpha_i')$ is the trine state
with
\begin{equation}
\alpha_i' = \frac{\alpha \sin^2 \phi_i }{\alpha \sin^2 \phi_i
+ \frac{1}{2} (1-\alpha) \cos^2\phi_i}
\label{alpha-prime}
\end{equation}
and 
\begin{equation}
p_i'  = 3 p_i \left[ \alpha \sin^2  \phi + {\textstyle{\frac{1}{2}}}(1-\alpha) 
\cos^2 \phi \right]
\label{p-prime}
\end{equation}
is the probability that we observe this trine state, given that we started
with $T_0(\alpha)$.  Similar formulae hold for the trine states $T_1$ and
$T_2$.  
We compute that 
\begin{equation}
\sum_i p_i' \alpha_i' = \sum_i 3 p_i \alpha \sin^2 (\phi_i) = \alpha.
\end{equation}
Also notice that the first stage of this process, 
the partial measurement which applies the matrices
$\sqrt{p_i}\, M(\phi_i)$, reveals
no information about which of $T_0$, $T_1$, $T_2$ that
we started with.  Thus, by the
chain rule for classical Shannon information \cite{Cover}, 
the accessible information 
obtained by our two-stage
measurement is just the weighted average (the weights being
$p'_i$) of the maximum over $\theta$ of the
Shannon mutual information $I_{\alpha_i'}(\theta)$ between 
the outcome of the von Neumann 
measurement $V(\theta)$ and the trines $T(\alpha_i')$.  
By convexity, it suffices to use 
only two values of $\alpha_i'$ to obtain this maximum.  In fact, the optimum
is obtained using either
one or two values of $\alpha_i'$ depending on whether the function
\[
I_{\alpha'} = \max_\theta I_{\alpha'}(\theta)
\]
is concave or convex over the appropriate region.  In the remainder
of this section, we give the results of computing (numerically) the values 
of this function $I_{\alpha'}$, and we
show that for small enough $\alpha$ it is convex,
so that we need two values of $\alpha'$.  We will then show that obtaining
this maximum requires a POVM with six outcomes.

\begin{figure}[th]
\leavevmode
\epsfxsize=6.5in
\epsfbox{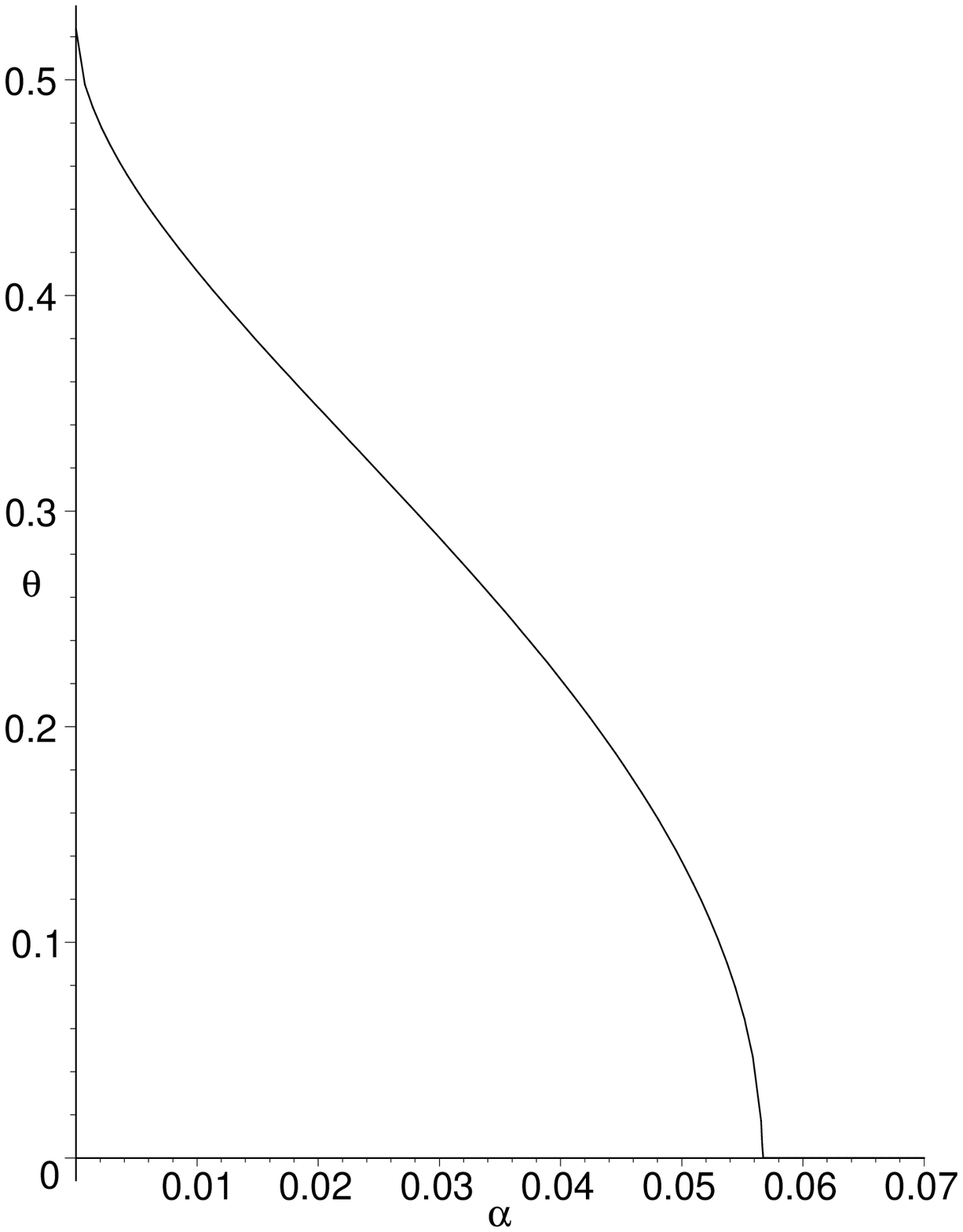}
\caption{The value of $\theta$ maximizing $I_\alpha$
for $\alpha$ between $0$ and $.07$. This function starts at $\pi/6$ 
at $\alpha=0$, decreases until it hits $0$ at
$\alpha = .056651$ and stays at $0$ for larger $\alpha$.
\label{fig-opttheta} }
\end{figure}

\begin{figure}
\leavevmode
\epsfxsize=6.5in
\epsfbox{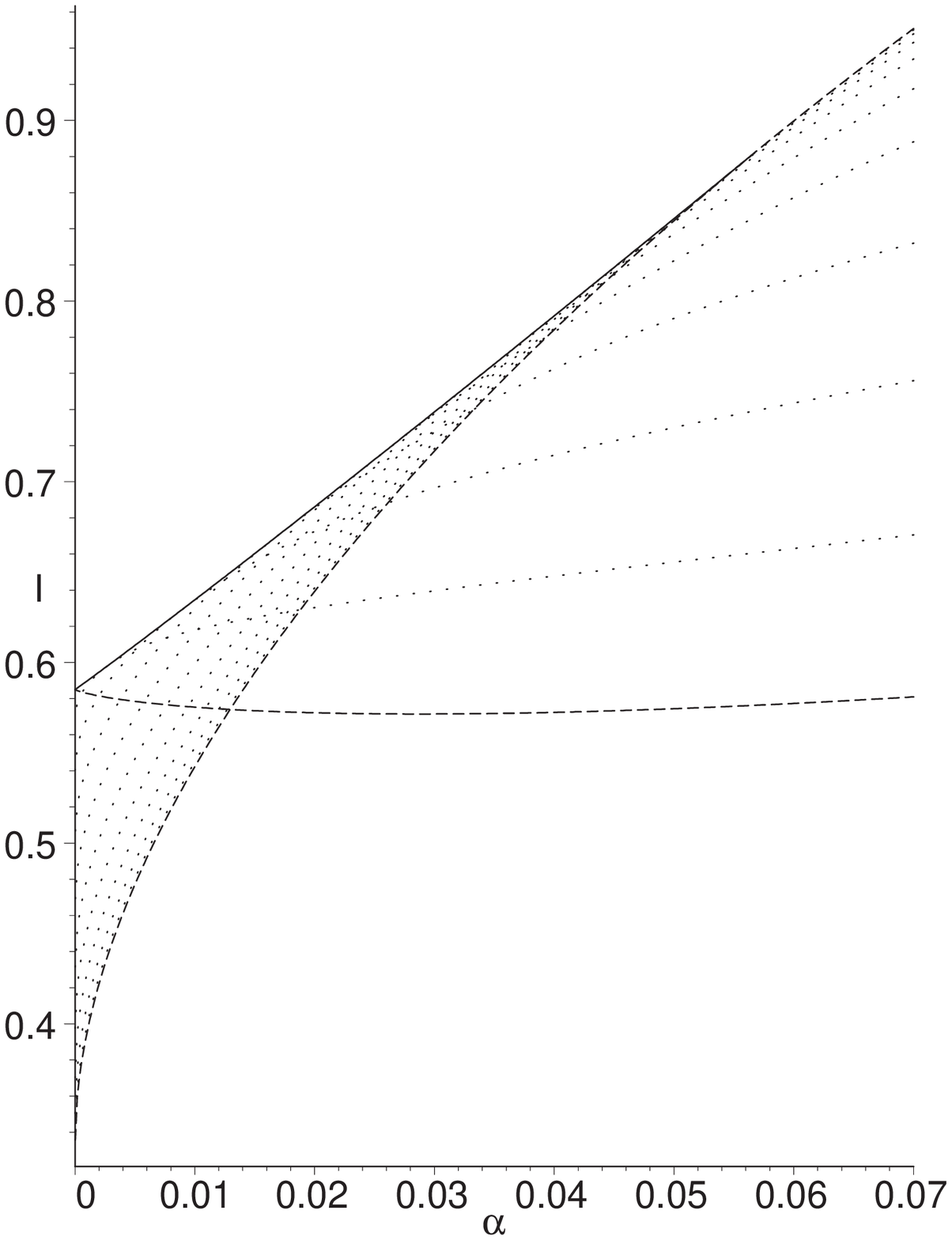}
\caption{This plot shows $I_\alpha(\theta)$ for 
various $\theta$.  The dashed curves are $I_\alpha(0)$
and $I_\alpha(\pi/6)$.  Note that $\theta=0$ is optimal for $\alpha > .056651$
and $\theta=\pi/6$ is optimal for $\alpha=0$.  The dotted curves show
$I_\alpha(\theta)$ for $\theta$ at intervals of $3^\circ$ between
$0$ and $\pi/6 = 30^\circ$.
The solid curve shows $I_\alpha(\theta_{\rm opt})$ for those 
$\alpha$ where neither $0$ nor $\pi/6$ is the optimal $\theta$.  
The solid curve is slightly convex;
this is clearer in Fig.~\protect{\ref{fig-accessible}}.
\label{fig-manythetas}
}
\end{figure}

We need to calculate the Shannon capacity of the channel whose
input is one of the three trine states $T(\alpha')$, and whose output is 
determined by the von Neumann measurement $V(\theta)$.  Because of
the symmetry, 
we can calculate this using only the first projector $V_0$.
The Shannon mutual information between the input and the output is 
$H(X_\mathrm{in}) - H(X_\mathrm{in} | X_\mathrm{out})$,
which is
\begin{equation}
I_{\alpha'} = \log_2 3 +\sum_{b=0}^2 
\braket{V_0(\theta)}{T_b(\alpha')}^2 
\log \left( \braket{V_0(\theta)}{T_b(\alpha')}^2 \right).
\end{equation}
We compute that
the $\theta$ giving the maximum $I_\alpha'$ is $\pi/6$ when
$\alpha' = 0$, decreases continuously to 0 at $\alpha' = .056651$ and
remains 0 for larger $\alpha'$.  (See Fig. \ref{fig-opttheta}.)  
This value $.056651$ corresponds to an angle
of .24032 radians ($13.769^\circ$).  This $\theta$ was determined
by using the computer package Maple 
to numerically find the point at which $d I_\alpha(\theta)/ d\theta = 0$.

\begin{figure}[th]
\leavevmode
\epsfxsize=6.5in
\epsfbox{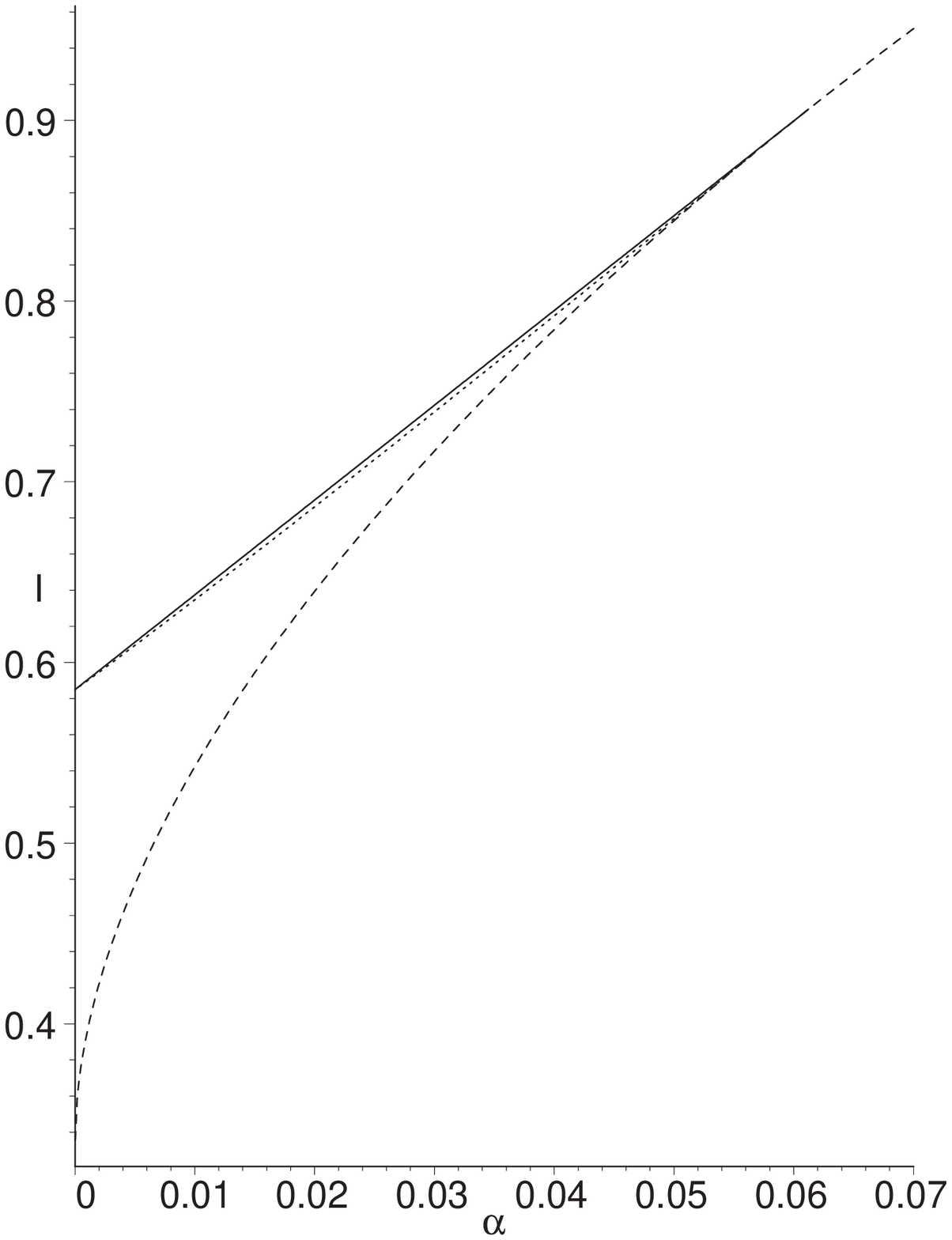}
\caption
{This graph contains three curves.  
The dashed curve is $I_\alpha(0)$ and the dotted curve is the maximum over
$\theta$ of $I_\alpha(\theta)$ for $\alpha < .056651$.
The solid curve is the convex envelope of the other two curves.
This solid curve is a
linear interpolation between $\alpha = 0$ and $\alpha = .061367$ 
and corresponds to a POVM having six elements.
It gives the accessible information for the lifted trine states $T(\alpha)$
when $0 \leq \alpha \leq .061367$.\label{fig-accessible}}
\end{figure}

By plugging this optimum $\theta$ into the formula for $I_{\alpha'}$, we 
obtain the optimum von Neumann measurement of the form $V$ above.  
We believe that this is also the optimal generic von Neumann measurement,
but we have not proved this.  
The maximum of $I_{\alpha'}(\theta)$ over $\theta$, and curves 
that show the behavior of $I_{\alpha'}(\theta)$
for constant $\theta$, are plotted in Fig.~\ref{fig-manythetas}.  
We can now observe that the first part of the curve is convex, and thus that
for small $\alpha$
the best POVM will have six projectors, corresponding to two values
of $\alpha'$.  We calculate that for trine states
with $\alpha < .061367$, the two values of $\alpha'$ giving the maximum
accessible information are
$0$ and $.061367$; we will let $\gamma_1 = .061367$ be this second value.
The trine states $T(\gamma_1)$ make an angle of
.25033 radians ($14.343^\circ$) with the $x$-$y$ plane.
The accessible information thus obtained is plotted in 
Fig.~\ref{fig-accessible}.

We can now invert the formula for $\alpha'$ (Eq.~\ref{alpha-prime}) 
to obtain a formula for
$\sin^2 (\phi)$, and substitute the value of 
$\alpha'= \gamma_1$ back into the formula 
to obtain the optimal POVM.  We find
\begin{eqnarray}
\nonumber
\sin^2(\phi_\alpha) &=& 
\frac{1-\alpha}{1+\alpha\left(\frac{2-3\gamma_1}{\gamma_1}\right)}\\
&=& \frac{1-\alpha}{1+29.591\alpha}
\label{phi-formula}
\end{eqnarray}
where $\gamma_1 = .061367$ as above.  
Thus, the elements in the optimal POVM we have found
for the trines $T(\alpha)$, 
when $\alpha < \gamma_1$,
are the six vectors
$P_b(\phi_\alpha, 0)$ and $P_b(0, \pi/6)$, where $\phi_\alpha$ is given by
Eq.~\ref{phi-formula} and $b= 0, 1, 2$.  
We must also prove there are no other POVM's which attain the same
accessible information.  
The argument above shows that any optimal POVM must
contain only projectors chosen from these six vectors:
only those two values of $\alpha'$ can
give the maximum capacity, and for each of these values of $\alpha'$ there 
are only three projectors in $V(\theta)$ which can
maximize $I_{\alpha'}$ for
these $\alpha'$.  It is easy to check that there is only one set of 
probabilities $p_i$ which make the above six vectors into a POVM, 
and that none of these probabilities are 0 for $0 < \alpha < \gamma_1$.  
Thus, for the lifted trine states with $0 < \alpha < .061367$, 
there is only one POVM maximizing accessible information, and it 
contains six elements, the maximum possible for real states by a
generalization of Davies' theorem \cite{Davies-gen}.

\end{document}